\documentclass
 [%
 reprint,
 superscriptaddress,
 showpacs,
 amsmath,amssymb,
 aps,
 prl,
 ]
 {revtex4-1}
 \usepackage{graphicx}
 \usepackage{dcolumn}
 \usepackage{bm}
 \usepackage{color}
 \usepackage{xcolor}
\begin{document}
%

\title{Parameter-free quantification of stochastic and chaotic signals}
\author{Sergio Roberto Lopes}\email{lopes@fisica.ufpr.br}
\affiliation{Universidade Federal do Paran\'a, Departamento de F\'{\i}sica, Curitiba, 81531-980, Brazil}
\author{Thiago de Lima Prado}
\affiliation{Universidade Federal dos Vales do Jequitinhonha e Mucuri, Instituto de Engenharia, Ci\^encia e Tecnologia,  Jana\'uba, 39440-146, Brazil}
\author{Gilberto Corso}
\affiliation{Universidade Federal do Rio Grande do Norte, Departamento de Biof\'isica e Farmacologia,  Natal, 59078-970, Brazil}
\author{Gustavo Zampier dos Santos Lima}
\affiliation{Universidade Federal do Rio Grande do Norte, Escola de Ci\^encias e Tecnologia, Natal, 59078-970, Brazil.}
\affiliation{Universidade Federal do Rio Grande do Norte, Departamento de Biof\'isica e Farmacologia,  Natal, 59078-970, Brazil}
\author{J\"urgen Kurths}
\affiliation{Potsdam Institute for Climate Impact Research - Telegraphenberg A 31,  Potsdam,  14473, Germany}
\affiliation{Humboldt University Berlin, Department of Physics, Berlin,12489, Germany}

\date{\today}

\begin{abstract}

Recurrence entropy $(\cal S)$ is a novel time series complexity quantifier based on recurrence microstates. Here we show that $\mathsf{max}(\cal S)$ is a \textit{parameter-free} quantifier of time correlation of stochastic and chaotic signals, at the same time that it evaluates property changes  of the probability distribution function (PDF) of the entire data set. 
$\mathsf{max}(\cal S)$ can distinguish   distinct temporal correlations of stochastic signals  following a power-law spectrum,  $\displaystyle P(f) \propto 1/f^\alpha$ even when shuffled versions of the signals are used. Such behavior is related to its ability to quantify  distinct subsets embedded in a time series. 
Applied to a deterministic system, the method brings new evidence about attractor properties and the degree of chaoticity. The development of a new parameter-free quantifier of stochastic and chaotic time series opens new perspectives to stochastic data and deterministic time series analyses 
and may find applications in many areas of science. 
\end{abstract}


\keywords{Recurrence entropy, stochastic signals, chaotic signals}
\maketitle

\section{Introduction}

Two of the foremost characteristics of a stochastic signal are its possible temporal correlation, preserving memory for some interval of time \cite{beran_2017} and the details of its probability distribution function (PDF), that bring information about how common can be an elements or a set of elements of a signal. Both characteristics are related to the concept of the  complexity of a signal, that we defined as a measure of how stochastic or dynamical systems express the degree of engagement of its elements in organized structured interactions.  High complexity is achieved in systems that exhibit a mixture of order and disorder and that have a high capacity to generate emergent phenomena, or in other words, the ability of a system as whole to display behaviors that can not be reduced to the properties of the constituent parts.  Despite the importance of the concept of signal complexity, no general and widely accepted means of measuring it  currently exists \cite{ref1,albert_2002}.

A common approach to characterize signal complexity is to use entropy-like quantities, describing the amount of data needed to identify the  state of a system \cite{shannon_1948}. 
Entropy is also a fundamental concept to understand chaotic dynamics \cite{kantz_2004} and can be related to the level of chaos or the chaoticity of the system,  mainly measured by the Lyapunov exponent \cite{kantz_2004,corso_2018}.   
Distinct time correlated stochastic signals are characterized by a frequency spectrum following a power-law distribution $\displaystyle P(f) \propto 1/f^\alpha$, where $\alpha$ quantifies the time correlation \cite{akaike_1974,beran_2017}. Specific values of $\alpha$ are associated with colors e.g. $\alpha=0$ for ``white", $\alpha=1$ for ``pink" or $\alpha=2$ for ``red" or, in this case also known as Brownian noise. 
Stochastic processes  with  $1/f^\alpha$ power spectra are ubiquitous in science finding applications in all its subareas like physics \cite{bak_1987,weissman_1988,press_1978,dos2012multifractality}, engineering  \cite{hooge_1981}, biology \cite{glass_2001,kobayashi_1982,west_1990,dos2014mouse}, cognition \cite{gilden_1995}, astrophysics \cite{press_1978, weissman_1988}, geophysics \cite{weissman_1988,matthaeus_1986}, economics \cite{granger_1996}, psychology \cite{gilden_2001}, language and music \cite{voss_19751}. 

A trustful method for estimation long-time correlation based on finite time series is key issue and, hitherto, an open question in time series analyses \cite{simonsen_1998, carbone_2007,weron_2002,podobnik_2008}. Many of these methods are based on time correlation quantifications such as the computation of the Hurst exponent \cite{carbone_2007,weron_2002}, detrended fluctuation analysis \cite{podobnik_2008} or range-scaled analysis \cite{weron_2002}. Others  are computed on the frequency or wavelet domains like periodogram or Wavelet methods \cite{simonsen_1998}.  The non-stationarity imposed by the long-range dependence ($\alpha>1$) associated to the finite time of the signal makes the characterization of correlation via those traditional methods a sophisticated technique. Often the analyses lead to parameter dependent results.  Empirical time series are always finite and long-range correlations are, unavoidably, partly suppressed. Diversely, the local dynamics characteristics of small temporal windows tend to be overestimated. 

On the other hand, the quantification of special properties of the PDF of a signal and its relation to the complexity  are also open questions. In fact, 
many attempts to quantify signal complexity have been developed in order to evaluate properties of the set of points composing a time series, usually employing the measure of entropy \cite{shannon_1948,pincus_1991,bandt2002,eroglu_2014}, but they do not evaluate time correlations or are parameter dependent. 

In this context, the evaluation of the recurrence entropy \cite{corso_2018} of a signal and our definition  of $\mathsf{max}(\cal S)$ show to be a powerful parameter-free tool to examine time series correlations. 
We show that the new approach can evaluate short and long time correlations, possesses a good agreement with traditional methods, and going further, providing   information about characteristics of the entire set of points of a signal.

The article is organized as follows: The recurrence entropy concept is introduced in section II, section III is devoted to the analyses and discussions  of time-correlated and non-correlated stochastic signals; section IV presents results and discussions of the  deterministic signal problem; our conclusions and final remarks are shown in section V.

\section{The recurrence entropy}

A visual tool to display recurrences of a $K$ length time series is defined as a $ K \times K$ binary matrix  \cite{marwan_2007}
\begin{equation}
\mathbf{R}_{ij}= \begin{cases} 1, & \mbox{if }  ||\mathbf{x}_i -  \mathbf{x}_j|| \leq \varepsilon    \\ 0, & \mbox{if }  ||\mathbf{x}_i -  \mathbf{x}_j || > \varepsilon \end{cases}\, i, j =1\cdots K,
\label{recurrece}
\end{equation}
where $\varepsilon$ is the vicinity parameter. $\mathbf{R}$ summarizes visually,  in a binary pattern, the information about how many recognizable subsets are embedded in a $K$ sequence of data showing how distinct will be the recurrence pattern (sequences of zeros and ones) of $K$ consecutive
points. The most explored subsets of $\mathbf{R}$ are diagonal lines of ``ones" representing the mutual recurrences of a sequence of points. However, other structures of $\mathbf{R}$ also have dynamical interpretations: the vertical/horizontal lines are associated to stationary points and the abundance of isolated points is an indicative of chaotic or stochastic dynamics  \cite{marwan_2007}. We generalize these concepts defining  recurrence microstates ${\cal A}(\varepsilon)$  as all possible cross-recurrence states among two randomly selected short sequences of $N$ consecutive points in a $K$ ($K\gg N$) length time series (we use $N=2,\; 3,$ and $4$), namely ${\cal A}(\varepsilon)$ are $N\times N$ small binary matrices. For example, supposing a time series of $K$ elements, $\{a_1,\, a_2,\,\cdots,\, a_{L-1},\, a_{L},\,a_{L+1},\,\cdots,\, a_K\}$ and using $N=2$, we randomly select two sequences of two elements, say $\{a_{10},\,a_{11}\}$ and $\{a_{L},\,a_{L+1}\}$. In the case of $N=2$  our microstates ${\cal A}(\varepsilon)$ will be  binary numbers composed of four elements ($0$ or $1$), namely a $2\times 2$ binary matrix expressing the cross-recurrences among $a_{10}$ and $a_L$,\, $a_{10}$, and $a_{L+1}$,\, $a_{11}$ and $a_{L}$ and, finally, $a_{11}$,\, and $a_{L+1}$. For a large enough randomly selected  number of samples $M$,  the recurrence entropy  $\cal S $ can be adequately computed by \cite{corso_2018} 
\begin{equation}
{\cal S}(\cal{A})=-\sum_{\cal A} P_{{\cal A}}  \;\textrm{ln}\; P_{{\cal A}}, 
\label{eq:entropy}
\end{equation}
where $P_{{\cal A}}$ measures the probability of occurrence of a specific state $\cal{ A}(\varepsilon)$ considering $M$ randomly samples.  Usually, $\varepsilon$ is a parameter-free as Eq. \ref{recurrece} and ${\cal A(\varepsilon})$  suggest, but this dependence is eliminated observing that $\cal S$ is null when computed for sufficient large or small $\varepsilon$, due to the absence of diversity of ${\cal A}(\varepsilon)$ for both cases. So, we impose a natural condition of a maximum for $\cal S(\varepsilon)$ \cite{jaynes_1957} turning  $\mathsf{max}(\cal S(\varepsilon))\equiv \mathsf{max}(\cal S)$ and $\cal A(\varepsilon)\equiv\cal A$ in  parameter-free quantities.

At first sight,  $M$ should be larger than the quantity of all possible microstates $2^{N^{2}}$, but as observed in \cite{corso_2018} the number of microstates effectively populated is small and the convergence of Eq. \ref{eq:entropy} is fast. So a much smaller number of randomly select microstates $M \sim 10,000$ is enough for good results in a large variety of cases and, in special, for all cases treated here, turning the method fast even for moderate values of microstate sizes $N$.

\section{Time correlated stochastic signals analyses}

\begin{figure}[htb]
\includegraphics[width=\columnwidth]{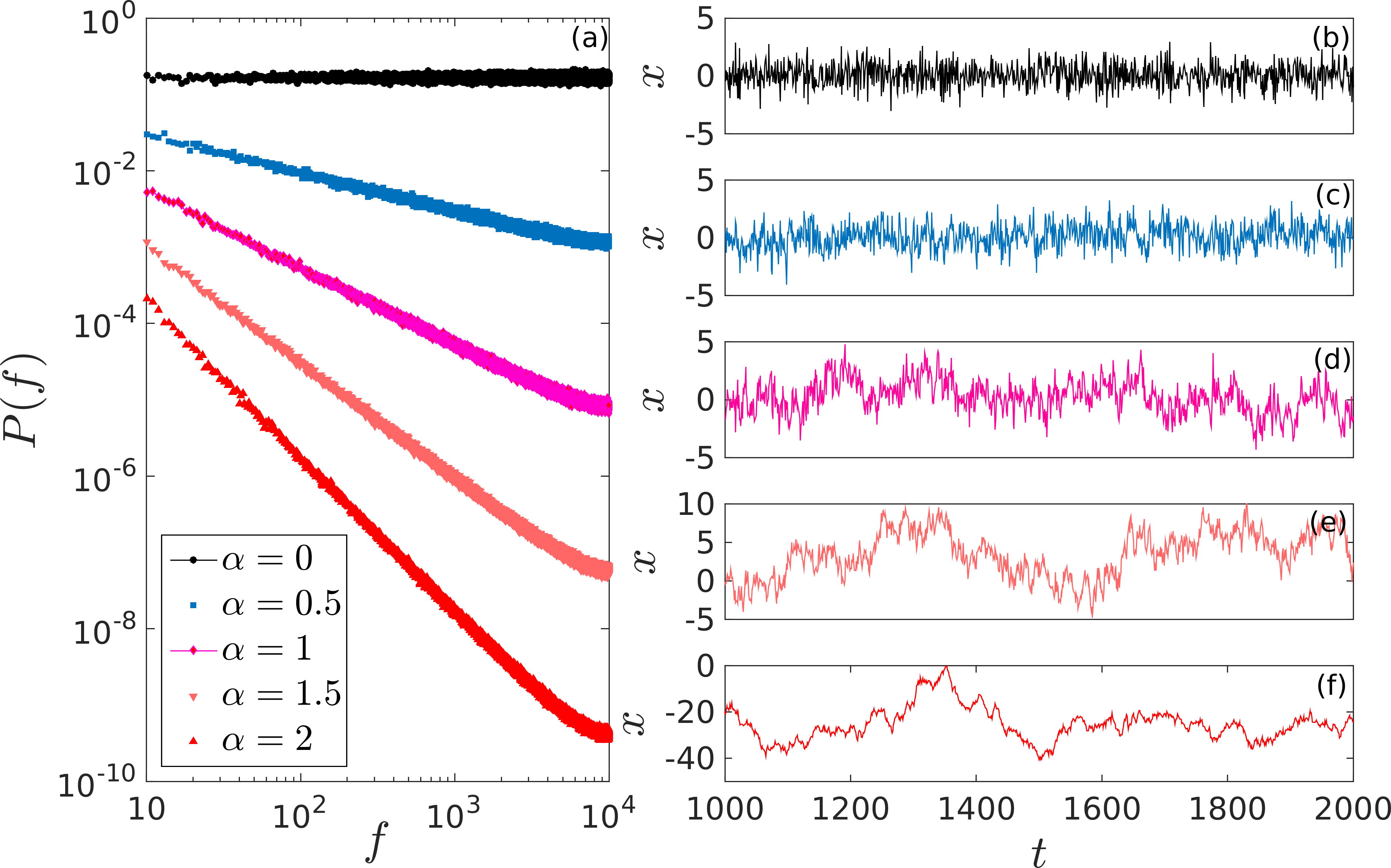}
\caption{\label{fig1} (a) Power spectral densities as a function of frequency ($\displaystyle P(f) \propto 1/f^\alpha$), for distinct $\alpha$ values. (b)--(f) Corresponding time series of the stochastic signals for all five values of $\alpha$. }
\end{figure}

Firstly, we consider time series of Gaussian distributed stochastic signals \cite{kasdin_1995}, characterized by a power spectrum $\displaystyle P(f) \propto 1/f^\alpha$ for $\alpha \in [0,\,2]$. Examples of the mean power spectrum obtained from $500$ time series are plotted in Fig. \ref{fig1}(a) for $5$ distinct values of $\alpha$. Corresponding individual time series examples are plotted in Figs. 1 (b-f).  
\begin{figure}[htb!]
\includegraphics[width=0.95\columnwidth]{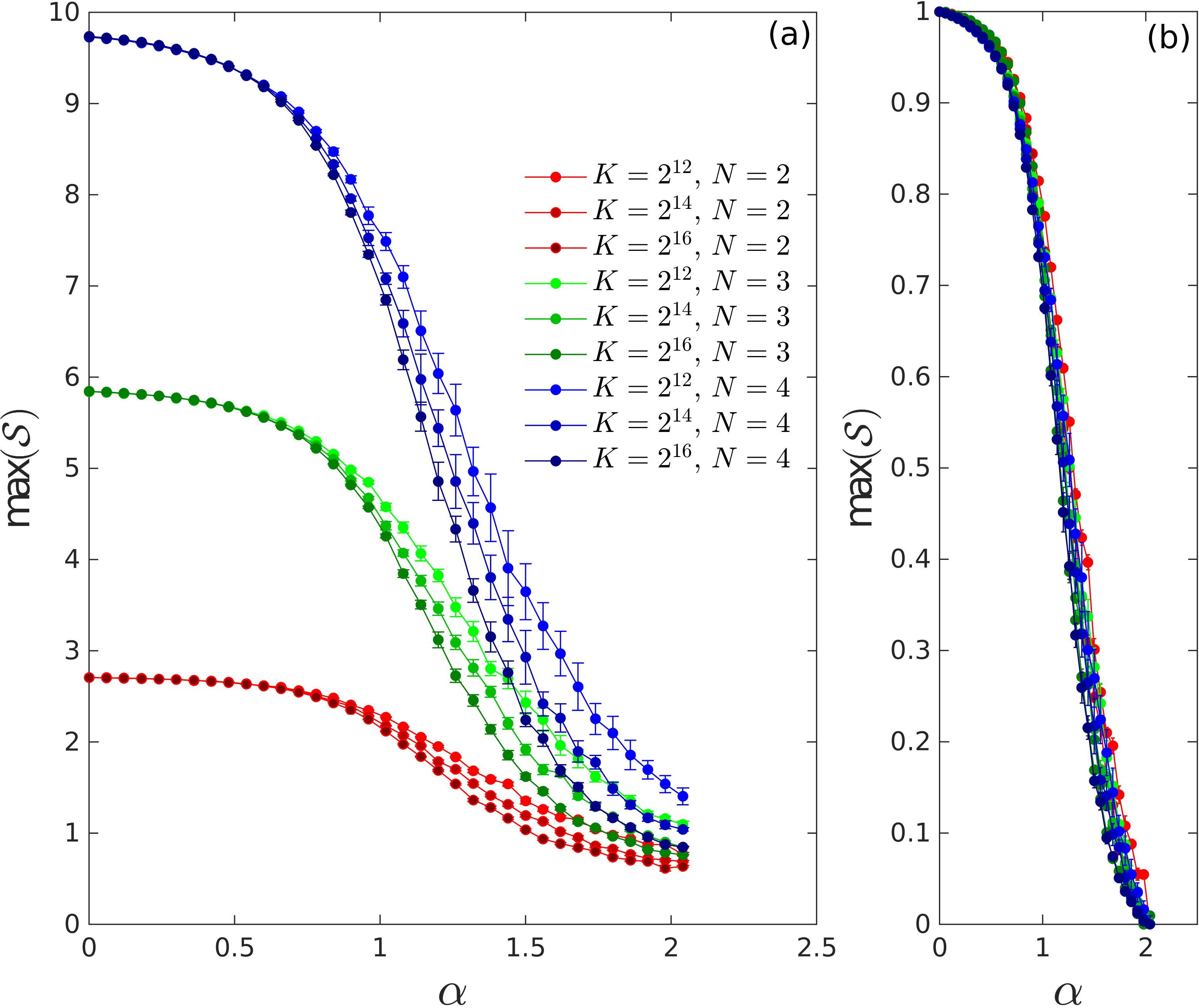}
\caption{\label{fig2} $\mathsf{max}(\cal S)$ as a temporal correlation quantifier. (a) $\mathsf{max}(\cal S)$ as a function of exponent $\alpha$ (stochastic time series shown in Fig.\ref{fig1}) for three  microstate sizes: $N=2,3$ and $4$ and distinct window sizes $K$. (b) The normalized $\mathsf{max}(\cal S)$ for all curves depicted in panel (a).}
\end{figure}

All properties of the stochastic signal are kept constant in the following analyses, but values of $\alpha \gtrsim 1.0$ impose a finite degree of non-stationarity due to long-term correlations as observed in Figs. \ref{fig1}(e, f).  For such cases, correlation-based methods overestimate (underestimate) short(long)-term correlations. Fig. \ref{fig2}(a) depicts the results of $\mathsf{max}(\cal S)$ computed for distinct colored stochastic signals ($0\leq \alpha\leq 2$) for $3$ values of $N$ and $3$ time series lengths. In general, $\mathsf{max}(\cal S)$ displays a typical logistic shaped curve as a function of $\alpha$. For vanishing values of $\alpha$, $\mathsf{max}(\cal S)$ will asymptote its maximum theoretical values $N^2\ln 2$, obtained for uncorrelated stochastic signals and infinite time series lengths. For the interval $0\leq \alpha\leq 2$, similar results for distinct $N$ show that the variability of $\mathsf{max}(\cal S)$ as a function of $\alpha$ is measurable even for the smallest possible value of the microstate matrix size $N=2$. Another important conclusion is that for a fixed $N$, longer time series lead to smaller values of $\mathsf{max}(\cal S)$ since longer time series provide a better evaluation of long-term time correlations. An error bar analysis specially for $N=4$ indicates that smaller time series associated to larger $\alpha$ and microstate size values result in larger dispersion of $\mathsf{max}(\cal S)$. This behavior reveals the natural dispersion expected for the quantification of long-term correlations when just finite time series are used. The results for $N=2$ and $N=3$ are less sensitive to the natural dispersion since the number of possible microstates are also smaller, such that tiny changes of the time correlation are not captured. All these features explored at the same time bring useful results when unknown source signals are analyzed.   Fig. \ref{fig2}(b) displays all curves depicted in Fig. \ref{fig1}(a) but normalized by its respective maximum. This data collapse reveals that the shape of  $\mathsf{max}(\cal S)$ for  all time series lengths and all microstate sizes are equivalent despite the small differences and details discussed above.
\begin{figure}[htb]
\includegraphics[width=\columnwidth]{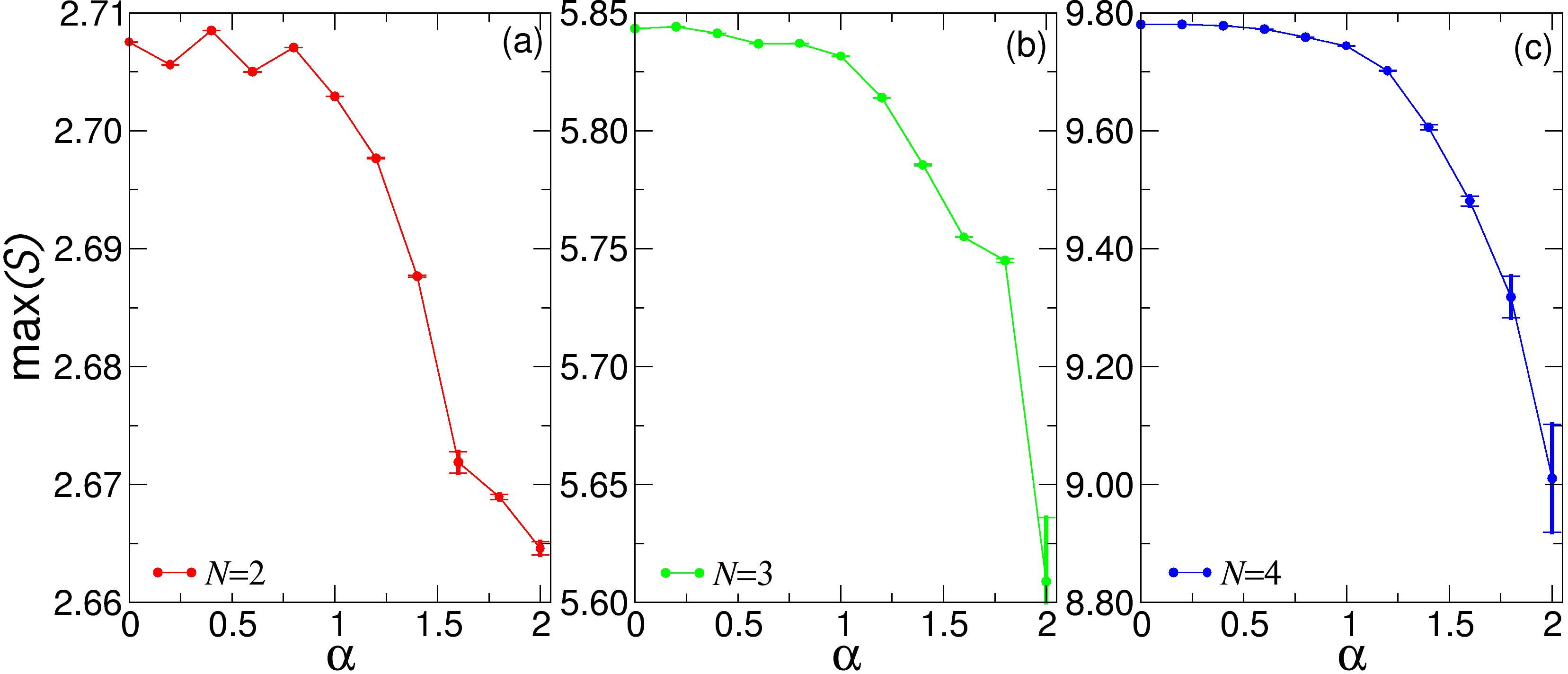}
\caption{\label{fig3} $\mathsf{max}(\cal S)$ of shuffled stochastic series (shown in Fig. \ref{fig1}) as a function of $\alpha$. We observed, even shuffled, a clear decay of absolute value of $\mathsf{max}(\cal S) $,  demonstrating its ability to distinguish distinct (even subtle) properties of the Gaussian PDFs. Error bars are due to $10$ numerical realizations.}
\end{figure}

Another important question about time series characterization is related to the characterization 
of the PDF of the signal.  To evaluate properties of the entire set of points in a time series, we make use of surrogate data analysis \cite{hair_2010}. One of the main and simple surrogate algorithm consists in shuffling the data, so that the data preserves the same amplitude distribution and mean, but  any correlation is destroyed, keeping only the collective properties of the set of points. For surrogate data methods the same analysis is carried out to the original data and the surrogated data to identify any distinguishable features between them. Traditional methods like Hurst exponents and detrended fluctuation analysis only quantify the time correlation \cite{beran_2017,podobnik_2008,simonsen_1998} and are not suitable for surrogated data.  

Fig. \ref{fig3} depicts the results of $\mathsf{max}(\cal S)$ applied to the same data used in Fig. \ref{fig2} for $K=2^{16}$ but shuffled in a random sequence (Fisher-Yates algorithm \cite{fisher_1963}) and using  $3$ values of $N$.  Now the results of $\mathsf{max}(\cal S)$ reveal a new question: even when the sequence of points in the time series is randomly organized, distinct stochastic signals lead to distinct values of $\mathsf{max}(\cal S)$. So the behavior of $\mathsf{max}(\cal S)$, in this case, is due only to properties of the set of points of the time series. Despite the fluctuation observed for $N=2$ and $N=3$, the results point out for a clear distinction between all our Gaussian PDFs of the time series. We observe that long-term correlations imposed by larger $\alpha$ results in a smaller value of $\mathsf{max}(\cal S)$ reflecting a more restrictive and more organized set of points, due to restrictions imposed by the correlation. Time correlations impose a limit to all possible sequence of subsets  in the time series and some combinations will not be allowed. 
\begin{figure}[htb]
\includegraphics[width=\columnwidth]{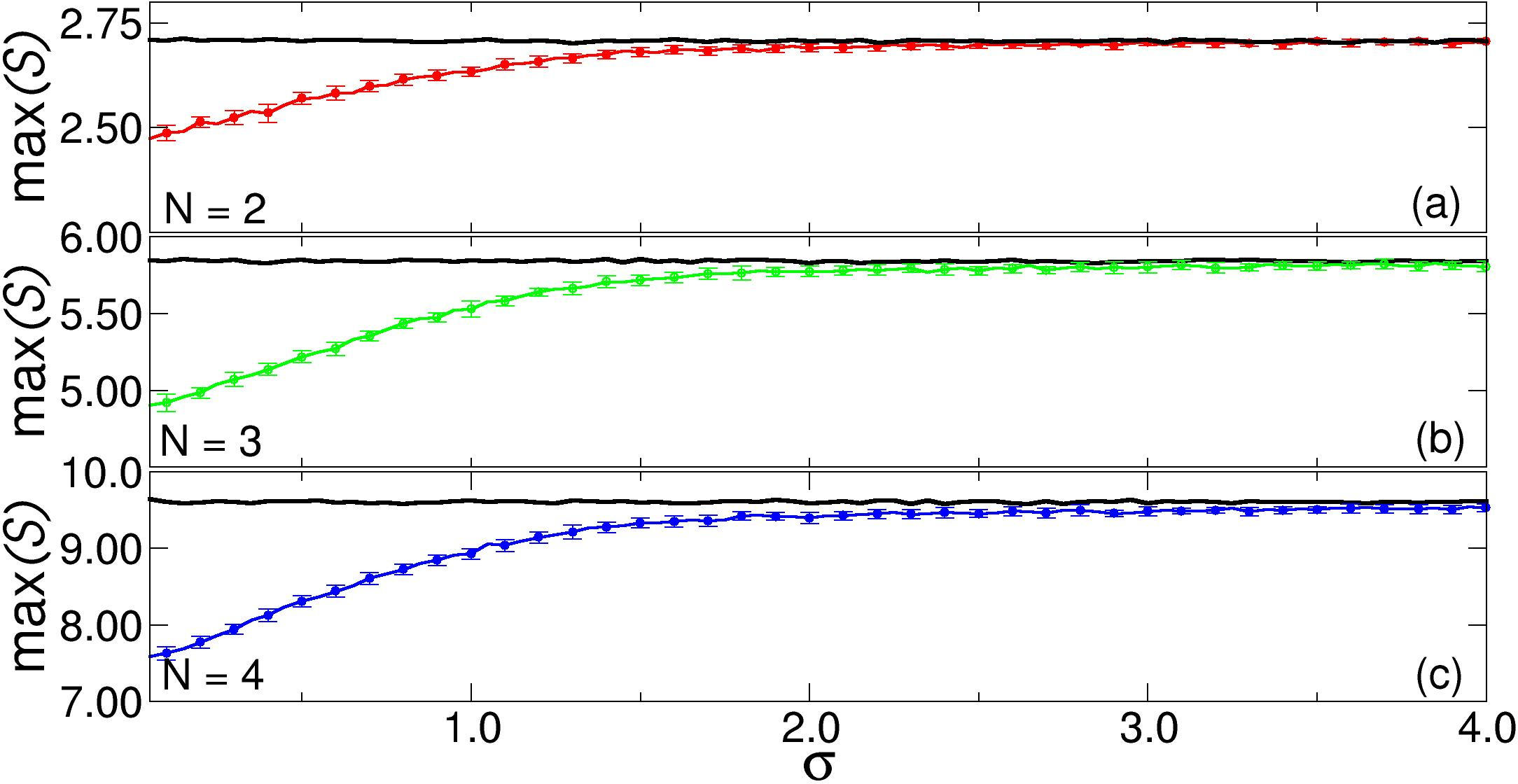}
\caption{\label{fig4} $\mathsf{max}(\cal S)$  of a shuffled sine with a uncorrelated (white) Gaussian noise added, as a function of the noise amplitude $\sigma$  (see Eq. \ref{seno_noise}). The monotonic increase of the absolute value of $\mathsf{max}(\cal S)$ reveals the ability of the method to differentiate distinct PDF of uncorrelated noise signals. The black lines in all panels show theoretical value of $\mathsf{max}(\cal S)$ for white Gaussian noise, for N=2, 3 and 4, respectively. Error bars are due to $100$ realizations.}
\end{figure}

To make this point clear, we analyze the results of $\mathsf{max}(\cal S)$ obtained for the time series produced by 
\begin{equation}
x(t)=\mathtt{shuffled}\,[\sin(t)]+\sigma\,\mathtt{rand}\,(t),
\label{seno_noise}
\end{equation}
where the $\mathtt{shuffle}$ process 
follows \cite{fisher_1963}, ``\texttt{rand}" is an uncorrelated Gaussian noise and $\sigma$ measures the level of uncorrelated noise superposed to the shuffled harmonic signal. Thus, the stochasticity  of this example comes from $2$ sources: the shuffled process in the sine signal and the random noise generator. Fig. \ref{fig4} depicts results of $\mathsf{max}(\cal S)$ as a function of the $\sigma$ for 3 values of $N$. The black lines in all panels indicate  $\mathsf{max}(\cal S)$ obtained for uncorrelated stochastic signal using the same time series length. For $\sigma =0$, the signal is an uncorrelated set of points, but its set of points is very restrictive, namely those points obtained from the function $\sin(t)$. In  this case, $\mathsf{max}(\cal S)$ is  consistently smaller than those ones expected for uncorrelated stochastic signal. For $0< \sigma < 2$, $\mathsf{max}(\cal S)$ grows monotonically, pointing out for an increasing number of distinct recurrence entropy microstates of the data set since the stochastic perturbation amplitude is being increased. For $\sigma>2$, the  uncorrelated stochasticity is large enough to turn the time series in an uncorrelated stochastic time series and the PDF will also reflect this situation. Consequently,  $\mathsf{max}(\cal S)$ asymptotically reaches the expected value for uncorrelated noise. So, $\mathsf{max}(\cal S)$ captures progressive increases of the complexity (measured by an increasing number of distinct microstates) imposed to the PDF  even when all analyzed time series are completely uncorrelated.  

\section{Deterministic signals analyses}

\begin{figure}[htb]
\includegraphics[width=\columnwidth]{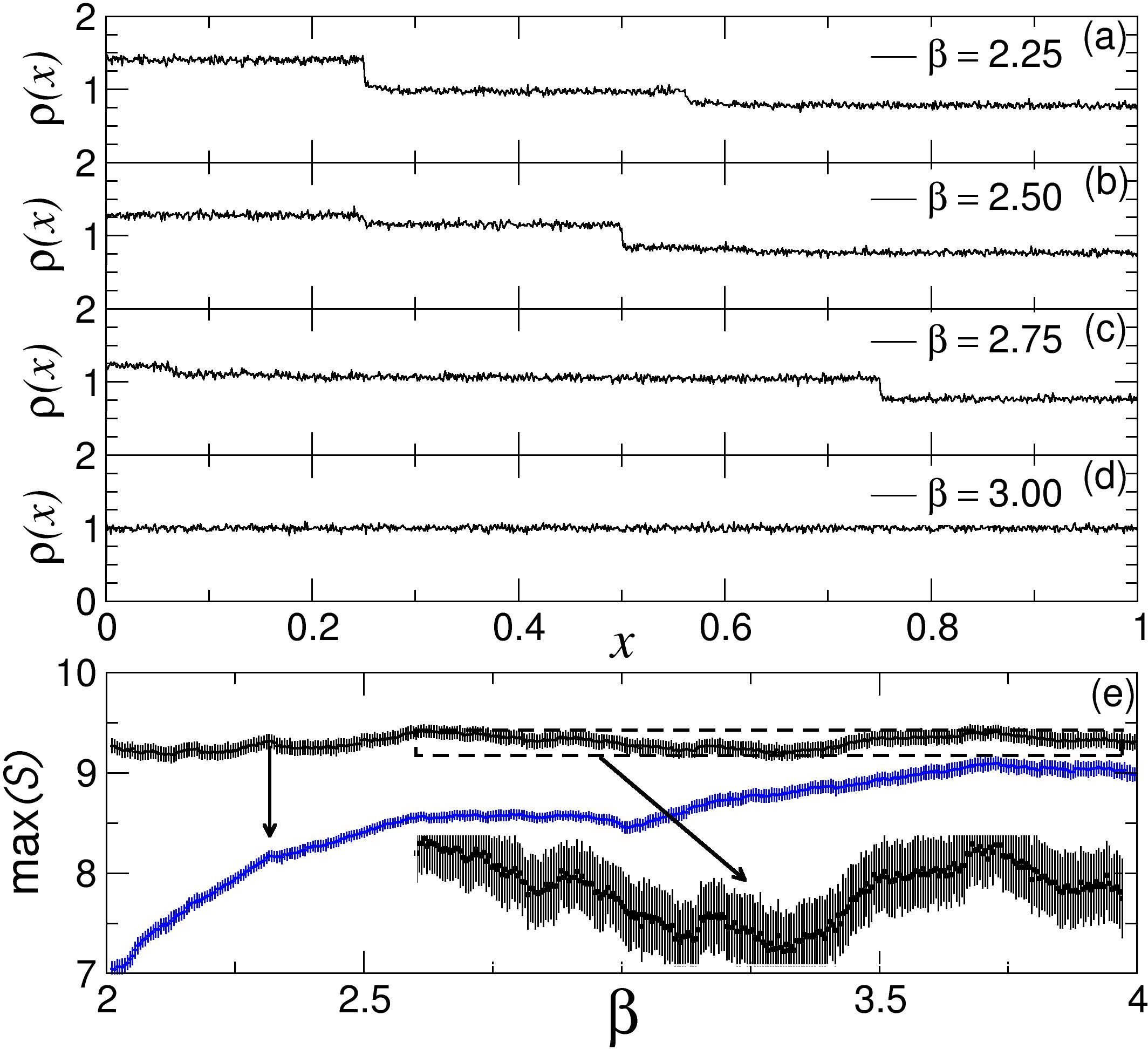}
\caption{\label{fig5} 
Panels (a-d) depict the invariant measure of the $x=\beta x\, \mathrm{(mod 1)}$ deterministic  map, for 4 distinct values of $\beta$. Panel (e) displays $\mathsf{max}(\cal S)$for time series  of the map, (blue curve) and shuffled time series (black). Notice that, as $\beta$ grows both chaoticity and the PDF complexity change. Absolute value of $\mathsf{max}(\cal S)$ (for $N=4$) reflects such changes in the PDF complexity level,  as observed in the shuffled time series of the map (black curve) (detailed in out of $y$ scale). $\mathsf{max}(\cal S)$ also reflects the increasing of chaoticity of the map,  imposed by the increase of $\beta$ (blue curve).}
\end{figure}

To prove the ability of $\mathsf{max}(\cal S)$ to capture distinct characteristics of even more complex PDFs, we analyze time series obtained by the generalized Bernoulli chaotic map
\begin{equation}
x=\beta\,x \hspace{1cm}(\mathrm{mod\, 1}).
\label{beta_x}
\end{equation}
For $\beta>1 $ the level of chaoticity can be evaluated by the Lyapunov exponent $\lambda = \ln\beta$ \cite{alligood_1996}. It is expected that entropy measures are related to $\lambda$ but not necessarily directly proportional since the entropy is also a function of the PDF of the attractor (the invariant measure) $\rho(x)$.  The quantity $\rho$ generated from Eq. \ref{beta_x} is  homogeneous for integer $\beta$, but becomes  inhomogeneous for non-integer values \cite{gora_2009}, due to the discontinuities observed in the PDF, result of an inhomogeneous measure of the attractor and a corresponding more complex signal. Figs. \ref{fig5}(a-d) display $\rho(x)$ of Eq. \ref{beta_x} map, depicting more complex PDFs for non-integer values of $\beta$ (a-c) but collapsing in a homogeneous one for integer $\beta$ (d). The entropy $\mathsf{max}(\cal S)$ as a function of $\beta$  will be a function of two factors, namely the continuous growing chaoticity associated to parameter $\beta$ superposed by a complex  behavior of the PDF.  Panel (e) shows the behavior of $\mathsf{max}(\cal S)$ in the interval $2<\beta<4$ (blue curve). The dashed tone of blue is representative of the standard deviation of $\mathsf{max}(\cal S)$ due to $100$ initial conditions for each $\beta$. To evaluate just the effect of changes in the PDF, the black curve in panel (e) depicts $\mathsf{max}(\cal S)$ computed for shuffled time series (out of $y$-scale magnification also shown). Again, the dashed black tone indicates the standard deviation over $100$ initial conditions. In this case, $\mathsf{max}(\cal S)$ depicts a complex oscillatory pattern due to the behavior of the PDF as discussed.  In general $\mathsf{max}(\cal S)$ grows as a function of $\beta$. However, the growth rate is faster for values of $\beta$ departing from integers, diminishing as $\beta$ approximates from the next subsequent integer.  Such behavior can be explained since for values immediately larger than each integer, the simultaneous increases of chaoticity and complexity of the PDF lead to a (local) maximum growth rate  of $\mathsf{max}(\cal S)$. As $\beta$ approximates to an integer, the complexity diminishes and, consequently, the rate of $\mathsf{max}(\cal S)$ also diminishes. As $\lambda$ increases with $\beta$, the chaoticity level increasing, in contrast to the complexity of the PDF that is decreasing, leading to smaller values of the growth rate of $\mathsf{max}(\cal S)$. For large values of $\beta$ as the interval $\beta \lesssim 4$ the growth rate can be even negative since a progressive less complex PDF can overtake the effect of the increasing of chaoticity. The value of $\mathsf{max}(\cal S)$ also reflects specific more dramatic changes of the PDF as the example highlighted by the arrow around $\beta \sim 2.3$, where a clear change in the PDF revealed by the shuffled time series analysis (black curve) leads to local small changes in the growth rate of $\mathsf{max}(\cal S)$ (blue curve). In resume the complex change behaviors of the PDF lead to a rich fine structure in $\mathsf{max}(\cal S)$ computed over the shuffled time series, denouncing a strong nonstationary time series (due to parameter changes in this example). 

\section{Conclusions}

In conclusion, we have shown that $\mathsf{max}(\cal S)$ is a parameter-free quantifier that additionally to the possibility to quantify time correlation of stochastic and chaotic signals, it goes further,  evaluating subtle properties of the PDF  of a signal, what can be computed using simple shuffling of the points. When time correlation is evaluated, $\mathsf{max}(\cal S)$ brings similar results to those obtained for more traditional but parameter-dependent quantifiers, such as the Hurst exponent. However the use of $\mathsf{max}(\cal S)$ makes clear a more complex interrelation about properties of the PDF and the complexity of the time series, bringing  new perspective for stochastic and chaotic data analyses. Our results can be useful in the analysis of experimental noisy data such as seismic, paleontology, economic problems where the possibility to evaluate properties of the entire data set data associated with the quantification of time correlation are important.  

We have analyzed  stochastic and deterministic signals. For both cases we conclude that the new method identify and quantify a new cause/effect relation where changes occurring in the time series PDF can be related directly to variations of complex behavior including the possibility to display short and long time correlations.  Finally, it is worth to mention that due to its computation methodology \cite{corso_2018}, recurrence entropy is fast evaluated for arbitrary long real-world time series, leading to robust parameter-free way to process data.

This study was financed in part by the Coordena\c c\~ao de Aperfeiçoamento de Pessoal de N\'{\i}vel Superior - Brazil (CAPES) - Finance Code 001 and trough project number  88881.119252/2016-01, Conselho Nacional de Desenvolvimento Cient\'{\i}fico e Tecnol\'ogico,  CNPq - Brazil, grant number 302785/2017-5, 
and Financiadora de Estudos e Projetos (FINEP).

\bibliography{biblio}

\end{document}